\documentclass[prl,twocolumn,superscriptaddress]{revtex4-2}
\usepackage{}
\usepackage{amssymb}
\usepackage{amsfonts}
\usepackage{bbm}
\usepackage{mathrsfs}
\usepackage{graphics,graphicx,epsfig,bm,amsmath,amsthm,amssymb}
\usepackage{bm}
\usepackage{bbm}
\usepackage{longtable}
\usepackage{multirow}
\usepackage{array}
\usepackage{color}
\usepackage[usenames,dvipsnames]{xcolor}
\usepackage{float}
\usepackage{textcomp}
\usepackage{upgreek}

\usepackage[a4paper,colorlinks=true,
linkcolor=blue,citecolor=blue,
pdfauthor={ },
pdftitle={ },
pdfsubject={ },
pdfkeywords={ }]{hyperref}

\begin{document}
\bibliographystyle{revtex4-2}

\title{Phased Geometric Controls of V-Shaped Three-Level System for Zero-field Quantum Sensing}

\author{Zhijie Li}
\thanks{These authors contributed equally to this work.}
\affiliation{CAS Key Laboratory of Microscale Magnetic Resonance and School of Physical Sciences, University of Science and Technology of China, Hefei 230026, China}
\affiliation{CAS Center for Excellence in Quantum Information and Quantum Physics, University of Science and Technology of China, Hefei 230026, China}

\author{Xiangyu Ye}
\thanks{These authors contributed equally to this work.}
\affiliation{CAS Key Laboratory of Microscale Magnetic Resonance and School of Physical Sciences, University of Science and Technology of China, Hefei 230026, China}
\affiliation{CAS Center for Excellence in Quantum Information and Quantum Physics, University of Science and Technology of China, Hefei 230026, China}

\author{Xi Kong}
\email{kongxi@nju.edu.cn}
\affiliation{The State Key Laboratory of Solid State Microstructures and Department of Physics, Nanjing University, 210093 Nanjing, China}

\author{Tianyu Xie}
\affiliation{CAS Key Laboratory of Microscale Magnetic Resonance and School of Physical Sciences, University of Science and Technology of China, Hefei 230026, China}
\affiliation{CAS Center for Excellence in Quantum Information and Quantum Physics, University of Science and Technology of China, Hefei 230026, China}

\author{Zhiping Yang}
\affiliation{CAS Key Laboratory of Microscale Magnetic Resonance and School of Physical Sciences, University of Science and Technology of China, Hefei 230026, China}
\affiliation{CAS Center for Excellence in Quantum Information and Quantum Physics, University of Science and Technology of China, Hefei 230026, China}

\author{Pengju Zhao}
\affiliation{CAS Key Laboratory of Microscale Magnetic Resonance and School of Physical Sciences, University of Science and Technology of China, Hefei 230026, China}
\affiliation{CAS Center for Excellence in Quantum Information and Quantum Physics, University of Science and Technology of China, Hefei 230026, China}

\author{Ya Wang}
\affiliation{CAS Key Laboratory of Microscale Magnetic Resonance and School of Physical Sciences, University of Science and Technology of China, Hefei 230026, China}
\affiliation{CAS Center for Excellence in Quantum Information and Quantum Physics, University of Science and Technology of China, Hefei 230026, China}
\affiliation{Hefei National Laboratory, University of Science and Technology of China, Hefei 230088, China}

\author{Fazhan Shi}
\email{fzshi@ustc.edu.cn}
\affiliation{CAS Key Laboratory of Microscale Magnetic Resonance and School of Physical Sciences, University of Science and Technology of China, Hefei 230026, China}
\affiliation{CAS Center for Excellence in Quantum Information and Quantum Physics, University of Science and Technology of China, Hefei 230026, China}
\affiliation{Hefei National Laboratory, University of Science and Technology of China, Hefei 230088, China}
\affiliation{School of Biomedical Engineering and Suzhou Institute for Advanced Research, University of Science and Technology of China, Suzhou 215123, China}

\author{Jiangfeng Du}
\email{djf@ustc.edu.cn}
\affiliation{CAS Key Laboratory of Microscale Magnetic Resonance and School of Physical Sciences, University of Science and Technology of China, Hefei 230026, China}
\affiliation{CAS Center for Excellence in Quantum Information and Quantum Physics, University of Science and Technology of China, Hefei 230026, China}
\affiliation{Hefei National Laboratory, University of Science and Technology of China, Hefei 230088, China}
\affiliation{School of Physics, Zhejiang University, Hangzhou 310027, China}

\begin{abstract}

Here we propose and demonstrate a phased geometric control protocol for zero-field double quantum gates in a V-shaped three-level spin system. This method utilizes linearly polarized microwave pulses and exploits the geometric qubit properties to prevent state leakage.  By employing specific phased geometric controls, we realize a low-power multi-pulse zero-field sensing technique using single nitrogen-vacancy centers in diamond. Our method offers a novel approach to implement precise double quantum gate operations with an adaptable driving power, making it a valuable tool for zero-field spin-based quantum technology.

\end{abstract}

\maketitle
\setlength{\parskip}{0pt}

In recent years, quantum sensing techniques based on controllable quantum systems have seen significant development. One successful example is the nitrogen-vacancy (NV) center in diamond, which possesses numerous merits, including nanoscale size, biocompatibility, and long coherence time under ambient conditions \cite{barry_sensitivity_2020,dolde_electric-field_2011,kucsko_nanometre-scale_2013,pham_nmr_2016,barson_nanomechanical_2017}. Typically, solid-state quantum systems require a static external magnetic field to lift the degeneracy of their ground-state manifolds. However, the presence of an external magnetic field suppresses the anisotropic interactions within the target sample, resulting in the loss of anisotropic physical information and causing inhomogeneous spectral broadening. A well-known zero-field technology is the zero- to ultralow-field nuclear magnetic resonance (ZULF NMR) spectroscopy. This technique effectively mitigates the inhomogeneous broadening of the spectrum in heterogeneous environments by attenuating the broadening effects induced by magnetic susceptibility \cite{barskiy_zero-field_2019}. More zero-field scenarios can be found in the field of electromagnetic biology \cite{ahonen_122-channel_1993,fenici_clinical_2005} and in the research of ferromagnetic film magnetization \cite{wei_ultralow_2021}. In order to extend the zero-field condition to  solid-state quantum systems like NV centers, the implementation of high-fidelity quantum control for the three-level system (3LS) is imperative.

To address the near-degenerate quantum states in the absence of external fields, one approach is to employ circularly polarized microwave pulses \cite{zheng_zero-field_2019,lenz_magnetic_2021}. While this method is effective when using a few pulses, it is limited in its ability to utilize double quantum (DQ) transitions with a multi-pulse method, which is crucial for sensing weak AC signals. Recent works have paved the way for realizing dynamical decoupling (DD) with linearly polarized microwave pulses at zero field by manipulating the 3LS via an effective Raman coupling \cite{cerrillo_low_2021}. This method enables the utilization of high-power multiple pulses, leveraging the advantage of DQ transitions at zero field to offer a significantly broader sensing bandwidth and expanded sensitivity range. However, the effectiveness of this method is compromised by the occurrence of state leakage due to the contradiction between the unavoidable hyperfine non-degeneracy and the limited driving field strength \cite{sekiguchi_geometric_2016,sekiguchi_dynamical_2019}. Subsequently, sequences that counteract the effects of the non-degeneracy detuning were proposed \cite{vetter_zero-_2022}. However, these methods, while relaxing the requirements for a strong driving field, lack versatility in their operations. In this study, we propose a method that prevents state leakage with a weak driving field by leveraging the geometric properties of the dressed states. Through this approach, a collection of effective DQ rotation operations can be achieved. Furthermore, we demonstrate a zero-field quantum sensing scheme utilizing single NV centers based on the proposed method.

\begin{figure*}
    \centering
	\includegraphics[width=0.97 \linewidth]{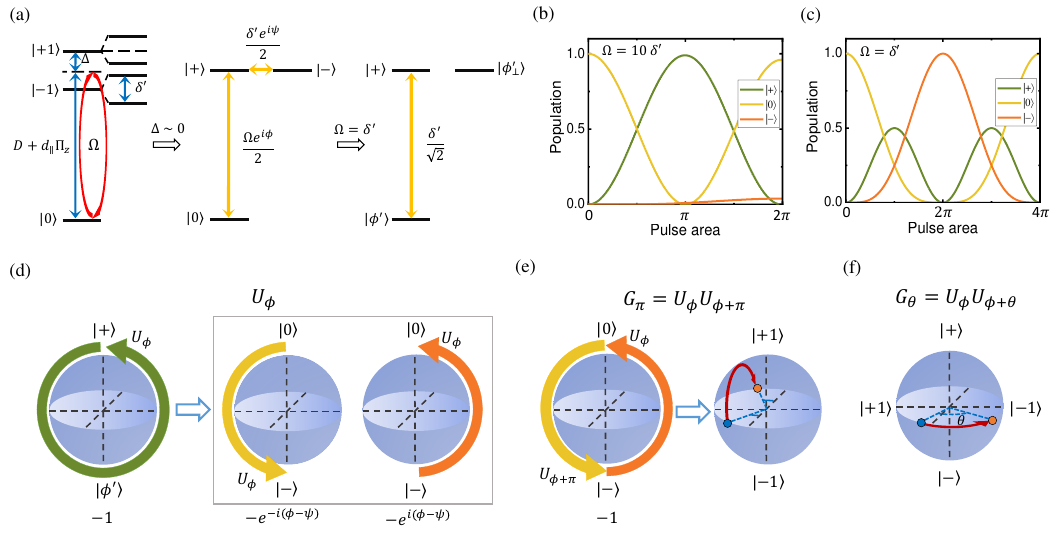}
    \caption{Illustration of phased geometric control protocol. (a) The spin triplet of the electronic ground state of NV centers. When the system is driven by a linearly polarized microwave pulse with amplitude $\Omega$, angular frequency $\omega=D+d_\parallel \Pi_z$ and phase $\phi$, Rabi oscillations $|0\rangle\leftrightarrow|-1\rangle$ and $|0\rangle\leftrightarrow|+1\rangle$ are activated simultaneously (left). Transformed to the $|0\rangle$, $|+\rangle=(|\!+\!1\rangle+|\!-\!1\rangle)/\sqrt2$ and $|-\rangle=(|\!+\!1\rangle-|\!-\!1\rangle)/\sqrt2$ basis, an effective Raman coupling is revealed (middle). When the driving power $\Omega$ is set to $\delta'$, the coupling only exists between $|+\rangle$ and $|\phi'\rangle$ states, where $|\phi'\rangle=(e^{i\phi }|0\rangle+e^{i\psi }|-\rangle)/\sqrt2$ is defined by the microwave phase $\phi$ and a coupling determined phase $\psi$, and $|\phi'_\perp\rangle=(e^{i\phi }|0\rangle-e^{i\psi }|-\rangle)/\sqrt2$ is the orthogonal counterpart to $|\phi'\rangle$ (right). (b) Evolution of the states driven with $\Omega=10\,\delta'$. The oscillation in the $\{|0\rangle,|+\rangle\}$ geometric qubit space dominates in the 3LS, while the transition to the state $|-\rangle$ is strongly suppressed in one cycle. (c) Evolution of the states driven with $\Omega=\delta'$. The complete transition between the states $|0\rangle$ and $|-\rangle$ can be realized, and the shortest pulse for a complete transition in the $\{|0\rangle,|-\rangle\}$ subspace corresponds to the $2\pi$ pulse in the $\{|+\rangle,|\phi'\rangle\}$ subspace. (d) Schematic diagram of the $2\pi$ operation $U_\phi$. The phase factors of final states brought by $U_\phi$ are shown respectively below each corresponding Bloch sphere. (e) Schematic diagram of the phased geometric gate operation $G_\pi=U_\phi U_{\phi+\pi}$. The state $|+\rangle$ undergoes a $4\pi$ cycle when $G_\pi$ is applied, which results in a phase factor valued $+1$. The $G_\pi$ operation brings a phase factor valued $-1$ in the $\{|0\rangle,|-\rangle\}$ subspace (left), and the effect of $G_\pi$ on the dynamical qubit is equivalent to that of a $\pi$ pulse (right), where the blue spot and the orange spot in the sphere represent the initial state and the final state respectively. (f) The effect of the operation $G_\theta=U_\phi U_{\phi+\theta}$. $G_\theta$ is equivalent to an effective rotation along $z$-axis in the $\{|+\rangle,|-\rangle\}$ subspace.}
    \label{Fig.1}
\end{figure*}

A single NV center in diamond consists of a substitutional nitrogen and a neighboring vacancy, its electron ground states form a typical 3LS (Fig.~\ref{Fig.1}(a)). The Hamiltonian of a single NV center driven by a linearly polarized microwave field can be given by ($\hbar=1$) \cite{vetter_zero-_2022,jamonneau_competition_2016}
\begin{align}
     H=&(D+d_\parallel \Pi_z) S_z^2+(\Delta+\delta/2)S_z+\Omega \mathrm{cos}(\omega t+\phi)S_x \nonumber \\
     &+d_\perp[\Pi_x(S_y^2-S_x^2)+\Pi_y(S_xS_y+S_yS_x)],
     \label{Hamiltonian}
\end{align}
where $\bm{S}=(S_x,S_y,S_z)$ is the spin-1 operator, $D$ is the zero-field splitting, $d_\parallel$ and $d_\perp$ are the longitudinal and transverse electric dipole moment components, $\Delta$ refers to the Zeeman splitting induced by the external magnetic field along the NV center's principle axis, $\delta$ contains hyperfine couplings with the surrounding spin-1/2 nuclei, and $\bm{\Pi}=(\Pi_x,\Pi_y,\Pi_z)$ denotes the total effective electric field. Furthermore, $\Omega,\omega$ and $\phi$ correspond to the amplitude, angular frequency, and phase of the linearly polarized microwave, respectively. Provided that the NV center's native nitrogen atom is a $^{15}\mathrm{N}$ atom and there is no magnetic field along the NV center's symmetric axis, the splitting within each electronic state manifold is primarily attributed to hyperfine interactions and transverse electric dipole couplings. When a linearly polarized microwave pulse with angular frequency $\omega=D+d_\parallel \Pi_z$ is applied, it drives the oscillations $|0\rangle\leftrightarrow|\!+\!1\rangle$ and $|0\rangle\leftrightarrow|\!-\!1\rangle$ simultaneously. As a result, an effective Raman coupling emerges (Fig.~\ref{Fig.1}(a)). By utilizing phase-fixed geometric controls \cite{sekiguchi_geometric_2016,sekiguchi_dynamical_2019,vetter_zero-_2022,cerrillo_low_2021} on the ground-state 3LS, it is possible to accumulate a geometric $\pi$ phase on the state $|+\rangle$ while keeping the state $|-\rangle$ nearly unchanged, as long as the $2\pi$ cycle occurs rapidly compared to the detuning modulation (Fig.~\ref{Fig.1}(b)). This approach enables the realization of a nearly $\pi$ pulse within the $\{|\!+\!1\rangle,|\!-\!1\rangle\}$ subspace. However, the presence of the hyperfine coupling $\delta$ and the transverse effective electric field $(\Pi_x,\Pi_y)$ can induce state leakage to the $|0\rangle$ state. Consequently, the imperfect controls in the dynamical decoupling sequence result in degraded spin coherence and distorted signal filtering, thereby diminishing the sensitivity.

In this Letter, we introduce a novel phased geometric control method that prevents state leakage and enables a diverse range of operations. With the resonance condition $\omega=D+d_\parallel \Pi_z$ and the microwave polarization perpendicular to the transverse projection of $\bm{\Pi}$, the Hamiltonian of the system can be expressed as \cite{supplemental_material}
\begin{equation} \label{Driven_Hamiltonian}
    \tilde{H'}(\Omega,\phi)=\left(\frac{\Omega e^{i\phi}}{2} |0\rangle + \frac{\delta'e^{i\psi}}{2} |-\rangle\right)\langle+|+\mathrm{H.c.},
\end{equation}
where $\delta'=\sqrt{\delta^2+4d^2_\perp\Pi_y^2}$ and $\psi=\arctan(-2d_\perp\Pi_y/\delta)$. Set $\Omega=\delta'$, a complete transition between the states $|0\rangle$ and $|-\rangle$ is activated (Fig.~\ref{Fig.1}(c)). The operation $U_\phi$, which enables the complete transition $|0\rangle\leftrightarrow|-\rangle$, is defined by the incident microwave phase $\phi$. Defining $|\phi'\rangle=(e^{i\phi }|0\rangle+e^{i\psi }|-\rangle)/\sqrt2$, the Hamiltonian Eq.~\eqref{Driven_Hamiltonian} can be written as
 \begin{equation}\label{Simplified_Hamiltonian}
    \tilde{H'}(\delta',\phi)=\frac{\delta'}{\sqrt{2}}\left(|\phi'\rangle\langle+|+|+\rangle\langle\phi'|\right).
\end{equation} 
In the qubit spanned by $\{|+\rangle,|\phi'\rangle\}$, Eq.~\eqref{Simplified_Hamiltonian} is proportional to the Pauli-X operator, and $U_\phi$ acts as a $2\pi$ pulse defined by the duration $T'=\sqrt{2}\pi/\delta'$ (Fig.~\ref{Fig.1}(c)). In this geometric spin qubit, any $2\pi$ cycle generates a microwave-phase independent factor of $-1$ before $|+\rangle$ \cite{sekiguchi_geometric_2016}. Moreover, the operation $U_\phi$ introduces conjugate phase factors in the $\{|0\rangle,|-\rangle\}$ subspace (Fig.~\ref{Fig.1}(d)), i.e. 
\begin{equation}
    \begin{aligned}
    &\langle -| U_\phi |0\rangle=-e^{-i(\phi-\psi)},\\
    &\langle 0| U_\phi |-\rangle=-e^{i(\phi-\psi)}.
    \end{aligned}
\end{equation}

\begin{figure}[tb]
    \centering
    \includegraphics[width=1 \linewidth]{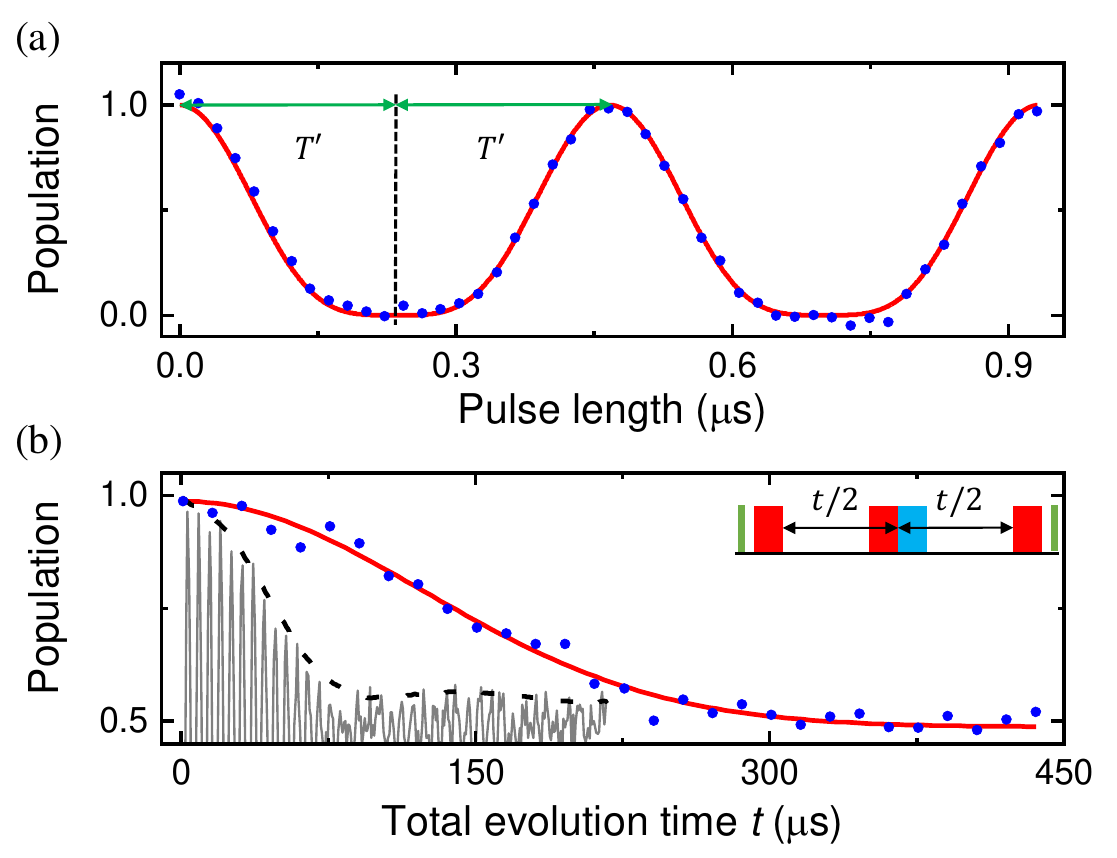}
    \caption{The state evolution under continuous driving field and pulsed experiments. (a) The population of the state $|0\rangle$ under the continuous driving field with different durations. The NV spin state is initialized and read out using $532\mathrm{\,nm}$ laser pulses. A microwave field with an angular frequency of $\omega=D+d_\parallel\Pi_z$ is applied. A complete transition to $|-\rangle$ can be realized when $\Omega=\delta'$ is met, and a microwave pulse with a duration of $T'=233.5\mathrm{\,ns}$ corresponds to a $2\pi$ pulse in the $\{|0\rangle,|+\rangle\}$ subspace. (b) Ramsey interference and spin echo. The Ramsey sequence is conducted with two separate $2\pi$ pulses. The observed data (gray line) exhibits an undersampled oscillation with a frequency of $\delta'/2\pi$. The envelope of the oscillation (black dashed line) is fitted using the function $\mathrm{cos}(\omega t)\mathrm{exp}[-(t/T_2^*)^p]$ and yields a dephasing time $T_2^*=120(2)\,$\textmu s $(p=1.8(1))$ in the $\{|\!+\!1\rangle,|\!-\!1\rangle\}$ basis. In the inset, the spin echo sequence is depicted, featuring a $4\pi$ pulse in the middle of the Ramsey sequence. This $4\pi$ pulse is composed of two $2\pi$ pulses, one phased $0$ (red rectangle) and the other phased $\pi$ (blue rectangle). The coherence of the echo sequence is evaluated using the function $\mathrm{exp}[-(t/T_2)^p]$ with $T_2=173(9)\,$\textmu s $(p=2.0(1))$ (red line).}
    \label{Fig.2}
\end{figure}

\begin{figure*}[tb]
    \centering
    \includegraphics[width=0.95 \linewidth]{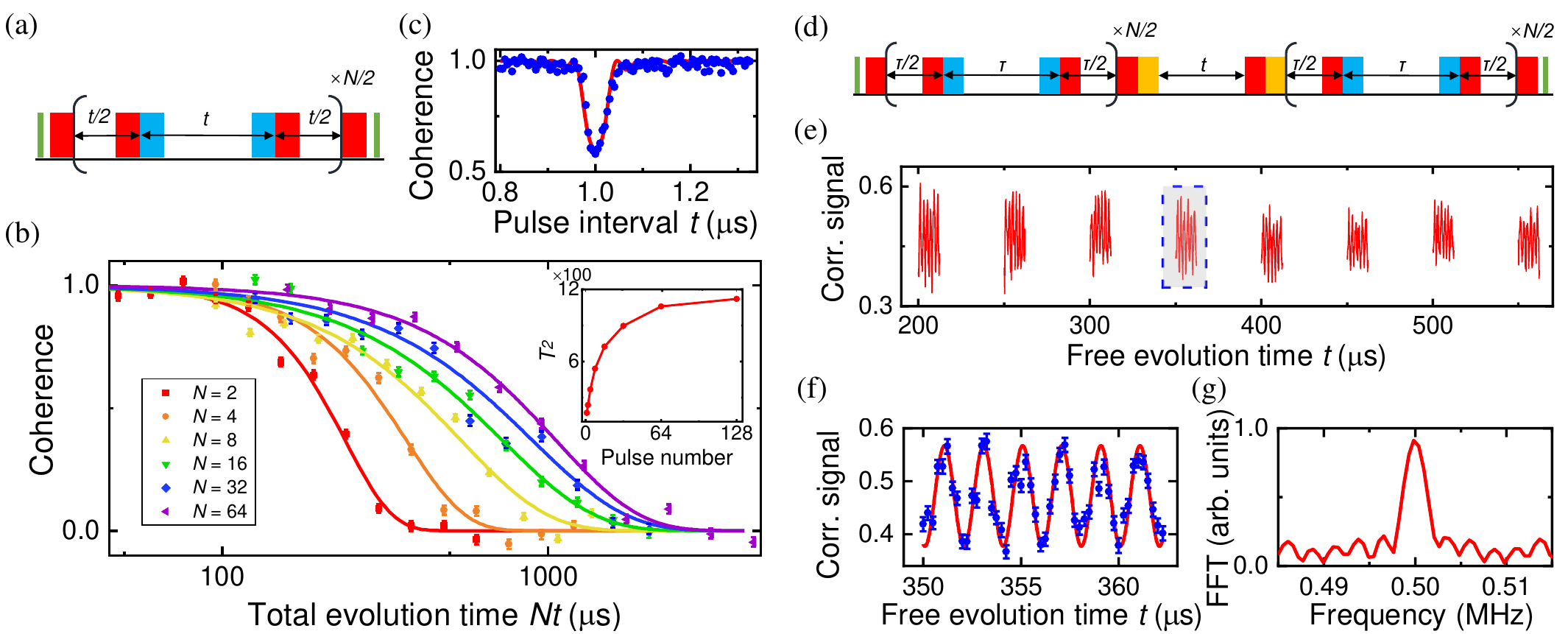}
    \caption{Decoherence measurements and correlation spectroscopy. (a) The ZDD-N sequence. The red and blue rectangles represent $U_0$ and $U_\pi$ operations respectively. (b) Decoherence measurements under ZDD-N sequences. The inset depicts the coherence saturation as the pulse number increases, ranging from the DQ spin echo to the ZDD-128 sequence. (c) AC signal measurement with the ZDD-64 sequence. The coherence dip is fitted by a fluctuating magnetic field of strength $B_{\mathrm{RMS}}=24.3(5)\mathrm{\,nT}$ with $t_s=1.002(1)\,$\textmu s. (d) The zero-field correlation spectroscopy sequence. The yellow rectangles represent the $U_{\pi/2}$ operations. (e) AC signal measurement with the correlation spectroscopy sequence. The phases accumulated during two ZDD-16 sequences with $\tau=t_s$ are correlated. (f) The data of the boxed gray area shown in (e). Correlation data (blue dots) are fitted by a $0.5\mathrm{\,MHz}$ sinusoidal oscillation (red line). (g) The discrete Fourier transform of the correlation data in (e).}
    \label{Fig.3}
\end{figure*}
\noindent
Therefore, the $4\pi$ pulse defined as $G_\pi=U_\phi U_{\phi+\pi}$ precisely leads to $|+\rangle\rightarrow|+\rangle$ and $|-\rangle\rightarrow-|-\rangle$ (Fig.~\ref{Fig.1}(e)). Consequently, the $\pi$ pulse in the $\{|\!+\!1\rangle,|\!-\!1\rangle\}$ subspace can be achieved without any leakage to the state $|0\rangle$, directly bringing about the zero-field dynamical decoupling (ZDD) sequence with equally spaced $G_\pi$ operations. Generally, phased geometric gate $G_\theta=U_\phi U_{\phi+\theta}$ is equivalent to the phase gate $P(\theta)$ in the $\{|+\rangle,|-\rangle\}$ subspace \cite{supplemental_material}, thus the effect of $G_\theta$ can be depicted as a rotation on the Bloch sphere (Fig.~\ref{Fig.1}(f)). Following the scheme outlined above, arbitrary effective rotations along $z$-axis in the $\{|+\rangle,|-\rangle\}$ subspace can be implemented. In addition to the $G_{\pm\pi}$ gates, the $G_{\pm\pi/2}$ gates are particularly relevant in quantum sensing protocols due to their ability to convert coherence into state population in the $\{|\!+\!1\rangle,|\!-\!1\rangle\}$ basis, which can be used to perform correlation of phases accumulated in separate DD sequences.

We use a $^{12}\mathrm{C}$ enriched diamond chip implanted with $40\mathrm{\,keV}$ $^{15}\mathrm{N}^+$ ions for our experiments. To counterbalance the geomagnetic field, a set of permanent magnets is employed, reducing the field strength to below $0.005\mathrm{\,mT}$. In this regime, we ensure that $\Delta/\delta<<1$, where $\delta$ is dominated by the intrinsic $^{15}\mathrm{N}$ hyperfine interaction $A_\parallel$. The transverse microwave polarization is aligned perpendicular to the transverse effective electric field vector, with the polarization direction along the $x$-axis. The resultant non-degenerate splitting is given by $\delta'=\sqrt{A_\parallel^2+4d^2_\perp\Pi_y^2}=2\pi \times3.04(1)\mathrm{\,MHz}$. Therefore, the manipulating microwave can be determined by $\Omega=\delta'$ and $\omega=D+d_\parallel\Pi_z=2\pi \times 2870.79(1) \mathrm{\,MHz}$. Setting $\phi=0$, the $|+\rangle\leftrightarrow|0'\rangle$ transition is driven with the angular frequency $\overline{\Omega}=\sqrt{\delta'^2+\Omega^2}=\sqrt2\delta'$, and the pulse length of the $2\pi$ operation $U_{\phi}$ is defined by $T'=2\pi/\overline{\Omega}$ (Fig.~\ref{Fig.2}(a)). Applying the Ramsey sequence with two separate $2\pi$ pulses, oscillation of the frequency $\delta'/2\pi$ emerges. The envelope of this oscillation directly reflects the dephasing occurring in the $\{|\!+\! 1\rangle,|\!-\! 1\rangle\}$ subspace. By inserting $G_\pi$ in the middle of the Ramsey sequence, coherence revival is realized (Fig.~\ref{Fig.2}(b)). With the specific $4\pi$ pulse available, we construct the ZDD-N sequence in the form of $2\pi \mbox{-} (t'/2 \mbox{-}4\pi\mbox{-} t' \mbox{-} \overline{4\pi} \mbox{-} t'/2)^{N/2} \mbox{-}2\pi$ (Fig.~\ref{Fig.3}(a)), where $t'=t-2T'$ is the duration of each free evolution, $t$ denotes the pulse interval, $Nt$ is the total evolution time, and the superscript indicates the interchange of the phases of constituent $2\pi$ pulses. This interlaced sequence is designed to compensate fidelity errors caused by pulse imperfections up to the second order \cite{souza_robust_2012,genov_arbitrarily_2017,vetter_zero-_2022}. By applying the ZDD-N sequences, significant prolongation of the DQ coherence in the $\{|\!+\! 1\rangle,|0\rangle,|\!-\! 1\rangle\}$ basis is observed as the pulse number $N$ increases (Fig.~\ref{Fig.3}(b)), indicating that there are sufficient manipulation fidelity and coherence resources available for quantum sensing purposes.

Measurements of an AC signal with a frequency of $f=0.5\mathrm{\,MHz}$ are shown in Fig.~\ref{Fig.3}(c, e). The ZDD-64 sensed frequency is $f'=1/(2t_s)=0.499(1)\mathrm{\, MHz}$, corresponding to the coherence dip at $t_s=1.002(1)\,$\textmu s (Fig.~\ref{Fig.3}(c)). In nanoscale NMR applications, the correlation spectroscopy sequence \cite{laraoui_high-resolution_2013,smits_two-dimensional_2019} is utilized to achieve high-resolution spectroscopy or to mitigate the effects of unwanted harmonics \cite{loretz_spurious_2015}. However, conventionally performing this free precession technique at zero field is challenging due to the incomplete manipulation of the 3LS. Nevertheless, it can be implemented by inserting $G_{\pi/2}$ gates between separate DD sequences (Fig.~\ref{Fig.3}(d)). The lowest order correlation reveals the signal frequency \cite{laraoui_high-resolution_2013}, as expressed by
\begin{equation}
    \langle\mathrm{sin\psi_1}\mathrm{sin\psi_2}\rangle\sim\mathrm{cos} (2\pi f(2\tau+t) ), 
\end{equation}
where $\tau$ is set to $t_s$ according to the coherence dip in the ZDD spectrum, $\psi_i$ is the phase accumulated during each individual ZDD sequence. The correlation signal of two ZDD-16 sequences for the AC field sensed in Fig.~\ref{Fig.3}(c) is shown in Fig.~\ref{Fig.3}(e).

\begin{figure}[tb]
    \centering
    \includegraphics[width=0.9 \linewidth]{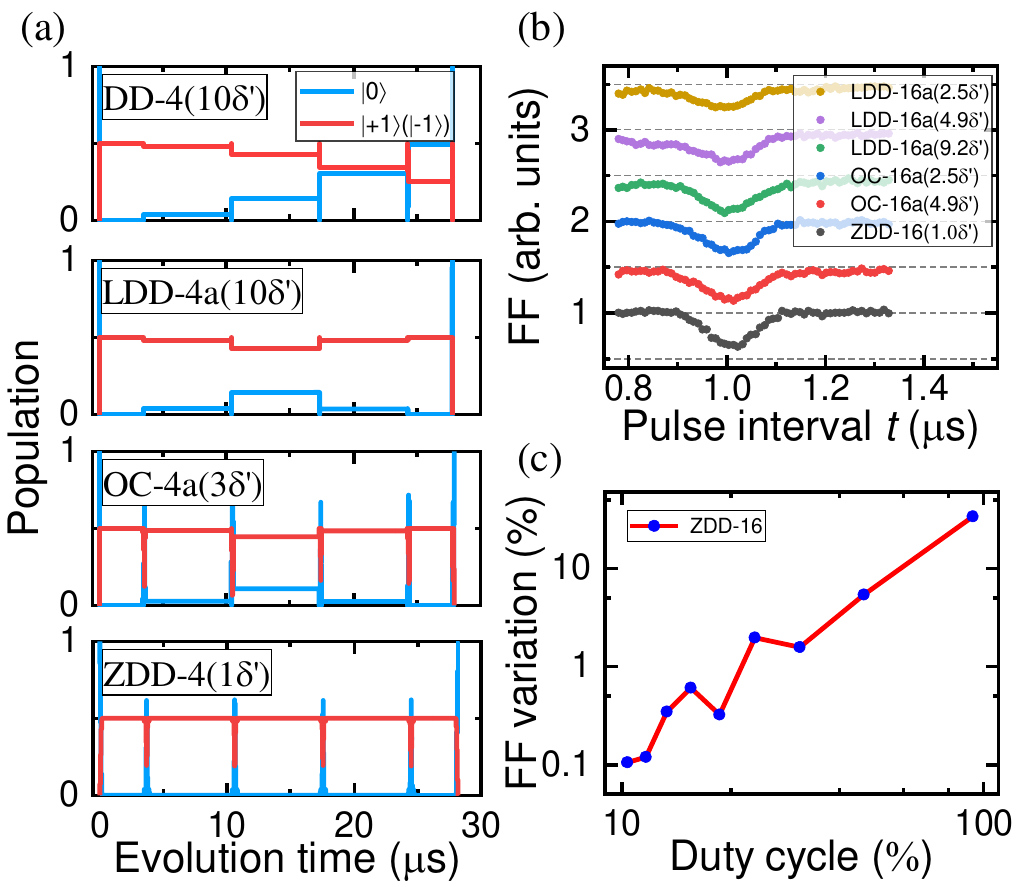}
    \caption{Comparison of ZDD and phase-fixed DD sequences. (a) Simulation of state evolutions under different DD sequences. (b) Measurements of the FFs at $\omega=0.5\mathrm{\,MHz}$. (c) Simulation of deviations for  the ZDD-16 FFs with respect to the ideal FFs under different duty cycles.}
    \label{Fig.4}
\end{figure}

In order to demonstrate the advantage of the ZDD sequence constructed with phased geometric gates, we conduct a comparison with other DD sequences. As shown in Fig.~\ref{Fig.4}(a), state evolutions of different DD sequences with distinct driving powers are simulated in the absence of signal fields. The state evolution under normal DD sequence is significantly distorted by detuning, while the LDD and the OC sequences \cite{vetter_zero-_2022} which utilize detuning-resistant phase arrangements as well as optimal control techniques, effectively suppress the distortion. In comparison, the ZDD sequence ensures equivalent populations during the free evolution periods.
Measurements of the filter functions (FFs) $F(t,\omega)$ of different DD sequences at $\omega=0.5 \mathrm{\, MHz}$ are presented in Fig.~\ref{Fig.4}(b). With low driving fields, the signal filtering of the LDD and the OC sequences are distorted. However, the ZDD sequence operating with $\Omega=\delta'$ exhibits a reasonable lineshape. The deviation between the ZDD-16 FF and the ideal FF is primarily caused by the finite duty cycle of the manipulating pulses. Nonetheless, this deviation is insignificant when the duty cycle is lower than $40\%$ (Fig.~\ref{Fig.4}(c)). In practice, the non-degenerate splitting $\delta'$ can be controlled by applying transverse strains, allowing for an adjustable duty cycle.

In this work, we introduce a phased geometric control protocol and demonstrate its application in a zero-field quantum sensing technique. The sequences employed for dynamical decoupling and correlation spectroscopy are specifically designed using phased geometric gates. Compared to previous approaches, our method provides a wider range of gate operations in sequence design and prevents the detrimental effects of state leakage by utilizing the properties of the geometric phase. In addition to the NV center, other solid spin systems such as divacancies in SiC \cite{koehl_room_2011} offer more alternatives for implementing the DQ manipulations with phased geometric gates. These systems possess a non-degenerate splitting that can be easily adjusted by strains or electric fields, enabling precise operations even with a short dephasing time. This allows for a broadened sensing bandwidth and the analysis of electric field noise. Furthermore, it is worth noting that our protocol can be extended to any other spin-based 3LS with similar energy configuration, thereby expanding its potential applications in various quantum technologies.

\section*{Acknowledgements}
This work was supported by the National Natural Science Foundation of China (Grant No. T2125011, 81788101), the National Key R\&D Program of China (Grant No. 2018YFA0306600), the CAS (Grant No. XDC07000000, GJJSTD20200001, Y201984, YSBR-068), Innovation Program for Quantum Science and Technology (Grant No. 2021ZD0302200, 2021ZD0303204), the Anhui Initiative in Quantum Information Technologies (Grant No. AHY050000), Hefei Comprehensive National Science Center, and the Fundamental Research Funds for the Central Universities.

This work was partially carried out at the USTC Center for Micro and Nanoscale Research and Fabrication. 

\bibliographystyle{unsrt}
\bibliography{ZDDcitation}
\end{document}